\documentstyle[12pt]{article}

\begin{document}

\begin{center}
{\bf Open cosmologies with rotation}
\end{center}

\begin{center}
Saulo Carneiro\footnote{Permanent address: Instituto de F\'{\i}sica,
Universidade Federal da Bahia, 40210-340, Salvador, Bahia,
Brazil.}
\end{center}

\begin{center}
{\it Instituto de Matem\'{a}ticas y F\'{\i}sica Fundamental \\ Consejo
Superior de Investigaciones Cient\'{\i}ficas \\ Serrano 121, 28006,
Madrid, Spain}
\end{center}

\begin{abstract}
We study a rotating and expanding, G\"{o}del type metric, originally
considered by Korotkii and Obukhov \cite{KO,Obukhov}, showing
that, in the limit of large times and nearby distances, it
reduces to the open metric of Friedmann. In the epochs when
radiation or dust matter dominate the energy density, our
solutions are similar to the isotropic ones and, in what
concerns processes occurring at small times, the rotation leads
only to higher order corrections. At large times, the solution
is dominated by a decaying positive cosmological term, with
negative pressure, and necessarily describes a quasi-flat
universe if the energy conditions have to be satisfied. The
absence of closed time-like curves requires a superior limit for
the global angular velocity, which appears as a natural
explanation for the observed smallness of the present rotation.
The conclusion is that the introduction of a global rotation, in
addition to be compatible with observation, can enrich the
standard model of the Universe, explaining issues like the
origin of galaxies rotation and the quasi-flatness problem.
\end{abstract}

\section{Introduction}

Around ten years ago, Korotkii and Obukhov presented a class of
rotating and expanding, G\"{o}del type cosmological metrics
\cite{KO,Obukhov}, showing that they respect the observed
isotropy of the cosmic background radiation and do not lead to
parallax effects. Furthermore, for some values of the metric
parameters, there are no closed time-like curves and, then,
these metrics do not suffer the causal problems characteristic
of the original G\"{o}del's metric \cite{Godel}.

In this paper we will show that, due to conservation of angular
momentum, the metric of Korotkii and Obukhov leads, in the limit
of large times, to an anisotropic metric that reduces to the
open metric of Friedmann in the nearby approximation. For small
times, we present an approximate solution valid in the limit of
small rotation, which presents an isotropic distribution of
pressures and the same evolution law as in the corresponding
isotropic case. Due to the rotation, the expressions for the
energy density and pressure are affected only by higher order
corrections relative to the standard, isotropic expressions,
which guarantees that anisotropy does not affect, unless by
higher order corrections, the processes occurred during early
times.

For the epoch dominated by dust matter, the corresponding
expanding solution of Einstein equations is similar to the
isotropic open solution, except for an anisotropic distribution
of pressures that, as we shall see, can be related to a material
content formed by an imperfect fluid. Nevertheless, the
anisotropy gives rise to an important difference in the later
stages of Universe evolution. For our solutions to satisfy the
dominant energy conditions, namely, positivity and causal flux
of energy, the epoch dominated by dust matter should naturally
be followed by an era of coasting evolution, in which the energy
density $\epsilon$ falls with $a^2$, where $a$ is the radius of
the Universe. This corresponds to a material content that
satisfies the equation of state $p=-\epsilon/3$. Such a content
can be interpreted as a decaying positive cosmological term, and
it is very significative that arguments from quantum cosmology
also predicts the conservation law $\epsilon a^2=$constant for a
time dependent cosmological term \cite{Wu}. Moreover, during
this phase, the relative energy density (the energy density
relative to the critical one) is a constant, and the energy
conditions impose a lower bound on its value which is close to
the present value. This constitutes a possible explanation for
the observed quasi-flatness of the Universe.

The general conclusion we will try to establish is that the
introduction of a global rotation into the Universe description,
in addition to agree with the observations that have been
sustaining the standard model, can shed light on subjects like
the origin of galaxies rotation, as pointed out by Li \cite{Li},
or the quasi-flatness problem.

\section{G\"{o}del type metrics}

The G\"{o}del type metric that we will consider is given by
\cite{KO,Obukhov,Saibatalov}
\begin{equation}
\label{KO}
ds^2=a^2(\eta)[(d\eta+le^xdy)^2-(dx^2+e^{2x}dy^2+dz^2)],
\end{equation}
where $a$ is a scale factor, $l$ is a positive parameter, $\eta$
is the conformal time and $x$,$y$,$z$ are spatial coordinates.

Korotkii and Obukhov \cite{KO,Obukhov} have shown that this
metric respects the observed isotropy of the cosmic background
radiation and does not lead to parallax effects, contrary to
what would be expected from an anisotropic, rotating metric.
Moreover, they have shown that, for $l<1$, there are no causal
problems, because the closed time-like curves characteristic of
G\"{o}del's metric can appear only for $l>1$ (the G\"{o}del metric
corresponds to $l=\sqrt{2}$, with $a$ constant\footnote{For an
exhaustive study of the stationary case of metric (\ref{KO}),
see the pioneer work of Rebou\c cas and Tiomno \cite{RT}.}).

Metric (\ref{KO}) describes an expanding and rotating universe,
with an angular velocity given, in comoving coordinates, by
$\omega=l/2a$ \cite{KO,Obukhov}. Although this result was
derived by Korotkii and Obukhov for a constant value of $l$, it
is easy to verify that it remains valid when $l$ is a function
of time. Using conservation of angular momentum, it is possible
to see that, in the radiation dominated epoch, the parameter $l$
is a constant, as originally considered by Korotkii and Obukhov,
while in the matter dominated one it falls with $a$. Indeed,
from the conservation of angular momentum, we have $\epsilon
\omega a^5$= constant, where $\epsilon$ is the energy density of
the matter content. In the radiation epoch, $\epsilon$ falls
with $a^4$, and so $\omega$ falls with $a$, leading to a
constant $l$. On the other hand, for the matter epoch $\epsilon$
falls with $a^3$, so $\omega$ falls with $a^2$, and $l$ should
fall with $a$. As we shall see, in a rotating and expanding
universe described by metric (\ref{KO}), the matter dominated
epoch should be followed by an era in which the energy density
falls with $a^2$, if the energy conditions have to be satisfied.
So, during this last epoch $\omega$ falls with $a^3$, and $l$
falls with $a^2$.

Therefore, for large times the terms in $l$ can be dismissed in
Einstein's equations, which means to consider, instead of metric
(\ref{KO}), the anisotropic metric
\begin{equation}
\label{aniso}
ds^2=a^2(\eta)[d\eta^2-(dx^2+e^{2x}dy^2+dz^2)].
\end{equation}
The cosmological solutions we will present in this paper,
approximate solutions of metric (\ref{KO}), are exact solutions
of the diagonal metric (\ref{aniso}), in the particular case
$l=0$.

With help of the coordinate transformation
\begin{eqnarray}
\label{coordenadas}
&e^x=\cosh{\xi}+\cos{\phi}\sinh{\xi}, \nonumber \\
&ye^x=\sin{\phi}\sinh{\xi},
\end{eqnarray}
metric (\ref{aniso}) can also be written as
\begin{equation}
\label{metrica}
ds^2=a^2(\eta)(d\eta^2-d\xi^2-\sinh^2\xi d\phi^2-dz^2).
\end{equation}
The coordinate transformation (\ref{coordenadas}) is a
particular case, for $l=0$, of a more general transformation,
with help of which the metric (\ref{KO}) can be expressed in
cylindrical coordinates \cite{Godel,RT,Radu}.

It is easy to show that, in the limit of nearby distances, that
is, up to subdominant terms in $\sinh\xi$, metric
(\ref{metrica}) reduces to the open FLRW metric. Indeed, using
the transformations
\begin{eqnarray}  \label{4}
&\sinh\xi=\sinh\chi\sin\theta, \nonumber \\
&z=\sinh\chi\cos\theta,
\end{eqnarray}
relating cylindrical and spherical coordinates, we obtain by
differentiation
\begin{eqnarray}
&\cosh\xi d\xi = \sinh\chi\cos\theta d\theta+\cosh\chi\sin\theta
d\chi, \nonumber \\ &dz=-\sinh\chi\sin\theta
d\theta+\cosh\chi\cos\theta d\chi.
\end{eqnarray}
So, by using
\begin{equation}
\frac{1}{\cosh^2\xi} = \frac{1}{1 + \sinh^2\xi}
\approx 1 - \sinh^2\xi =
1 - \sinh^2\chi\sin^2\theta,
\end{equation}
we have
\begin{equation}  \label{5}
dz^2+d\xi^2\approx\sinh^2\chi d\theta^2+d\chi^2.
\end{equation}
Finally, substituting (\ref{5}) and the first of equations
(\ref{4}) into (\ref{metrica}) leads to
\begin{equation}  \label{6}
ds^2\approx
a^2(\eta)[d\eta^2-d\chi^2-\sinh^2\chi(d\theta^2+\sin^2\theta
d\phi^2)],
\end{equation}
which is precisely the open FLRW metric in spherical coordinates
\cite{Landau}.

The requirement of absence of closed time-like curves can be
used to understand why the present observed superior limit for
the global angular velocity is so small. As said before, the
rotation parameter $l$ is given by $l=2\omega a$
\cite{KO,Obukhov}. Then, the causality condition $l<1$ applied
to the radiation dominated era implies that
$\omega_d<1/2a_d\approx 2.5
\times 10^{-15}s^{-1}$, where $\omega_d$ is the global angular
velocity at the time of decoupling between matter and radiation,
and $a_d$ is the radius of the Universe at that time,
$a_d\approx6\times 10^{22}$m \cite{Weinberg} (for a present
radius of the Universe given by $a\approx 1/H\sim10^{26}$m,
where $H$ is the Hubble parameter). As we have seen, during the
matter dominated era $\omega a^2=$constant, which leads to an
upper limit for the present angular velocity of matter given by
$\omega=\omega_d a_d^2/a^2\sim10^{-21}s^{-1}$, while for
radiation we obtain, from $\omega_{rad} a=$constant, the upper
limit $\omega_{rad}=\omega_d a_d/a\approx 1.5
\times 10^{-18}s^{-1}$. In this way, the absence of closed
time-like curves appears as a natural explanation for the
smallness of the present rotation.

\section{The radiation dominated era}

>From metric (\ref{KO}) and considering $l$ as a function of
time, we obtain the Einstein equations
\begin{equation}
\label{E'0}
\epsilon a^4 =
-\left(1-\frac{3l^2}{4}\right)a^2+3(1-l^2)\dot{a}^2
-2l\dot{l}a\dot{a},
\end{equation}
\begin{equation}
\label{E'1}
p_1a^4 =
\left( \frac{l^2}{4}+\dot{l}^2+l\ddot{l} \right) a^2
+(1-l^2)\dot{a}^2-2(1-l^2)a\ddot{a}+4l\dot{l}a\dot{a},
\end{equation}
\begin{equation}
\label{E'2}
p_2a^4 =
\frac{l^2}{4}a^2+(1-l^2)\dot{a}^2-2(1-l^2)a\ddot{a}
+2l\dot{l}a\dot{a},
\end{equation}
\begin{equation}
\label{E'3}
p_3a^4 =
\left( 1-\frac{l^2}{4}+\dot{l}^2+l\ddot{l} \right) a^2
+(1-l^2)\dot{a}^2-2(1-l^2)a\ddot{a}+4l\dot{l}a\dot{a}.
\end{equation}
Here, $\epsilon$ is the energy density, and $p_i$, $i=1,2,3$,
are the principal pressures. The dot means derivation with
respect to the conformal time. The other non-null components of
the energy-momentum tensor are proportional to $l$, which from
now on will be considered a small parameter, i.e., $l<<1$. In
this approximation, these non-diagonal components can be
neglected compared to the diagonal ones.

As suggested in section 2, let us adopt for the radiation
dominated era the ansatz $l=$constant. The above equations turn
out to be
\begin{equation}
\label{E0}
\epsilon a^4 =
-\left(1-\frac{3l^2}{4}\right)a^2+3(1-l^2)\dot{a}^2,
\end{equation}
\begin{equation}
\label{E1}
p_1a^4 = p_2a^4 =
\frac{l^2}{4}a^2+(1-l^2)\dot{a}^2-2(1-l^2)a\ddot{a},
\end{equation}
\begin{equation}
\label{E23}
p_3a^2=p_1a^2+1-\frac{l^2}{2}.
\end{equation}

Substituting in (\ref{E0}) the conservation law for radiation,
$\epsilon a^4 = a_0^2 =$ constant, and considering the limit
$a\rightarrow 0$, we obtain the solution
\begin{equation}
a=b\eta=\sqrt{2bt},
\end{equation}
where $t$ is the cosmological time, defined by $dt=a\;d\eta$,
and
\begin{equation}
b=\frac{a_0}{[3(1-l^2)]^{1/2}}.
\end{equation}
This is the same evolution law obtained by the isotropic model
in the limit of small times \cite{Landau}.

For the energy density, we then have
\begin{equation}
\label{radiation}
\epsilon=\frac{a_0^2}{a^4}=\frac{3}{4t^2}\left(1-l^2
\right),
\end{equation}
while it follows, from equations (\ref{E1}) and (\ref{E23}), in
the same limit $a\rightarrow 0$,
\begin{equation}
\label{pressoes}
p_i=p=\frac{\epsilon}{3},
\end{equation}
$i=1,2,3$, i.e., the equation of state for radiation, as
expected.

For $l<1$, equations (\ref{radiation}) and (\ref{pressoes}) give
$\epsilon>0$ and $|p_i|<\epsilon$, that is, the energy
conditions \cite{Hawking} are satisfied. For $l^2<<1$, those
equations give the same predictions as the isotropic model
\cite{Landau}. Another remarkable point is that, although the
distribution of pressures is in general anisotropic, in the
limit of small times we obtain an isotropic pressure. In
addition, the Hubble parameter at this epoch has the same time
dependence as in the standard model, namely $H=1/2t$, which
leads to the same ratio between the interaction rate and the
expansion one. The conclusion is that the thermal history of
this universe is the same predicted by the standard model. In
what concerns processes occurring during the initial stages of
Universe evolution, the anisotropy (and the rotation) may be
manifest only as higher order corrections.

\section{Cosmological solutions for large times}

As pressure and energy density decrease, the radiation dominated
era evolves until matter and radiation decouple from each other
and one enters a matter dominated epoch characterized by the
conservation law
\begin{eqnarray}
\label{intermediate}
&\epsilon a^3=2a_1,
\end{eqnarray}
where $a_1$ is a constant.

Adopting the ansatz $la=$constant, taking the limit of large $a$
and keeping only the dominant terms, the Einstein equations
(\ref{E'0})-(\ref{E'3}) reduce to
\begin{equation}
\label{E''0}
\epsilon a^3 =
-a + \frac{3\dot{a}^2}{a},
\end{equation}
\begin{equation}
\label{E''1}
p_1a^3 = p_2a^3 =
\frac{\dot{a}^2}
{a}-2 \ddot{a},
\end{equation}
\begin{equation}
\label{E''3}
p_3a^2 = p_1a^2 + 1.
\end{equation}

Substituting (\ref{intermediate}) into (\ref{E''0}), we have the
solution
\begin{eqnarray}
\label{intermediate2}
&a(\eta)=a_1\left[\cosh\left(\frac{\eta}{\sqrt{3}}\right)-1\right],
\end{eqnarray}
where we have absorbed an integration constant by a suitable
shift in the origin of conformal time $\eta$.

With this solution, the spatial Einstein equations
(\ref{E''1})-(\ref{E''3}) lead to the pressures
\begin{eqnarray}
\label{pI}
&p_1a^2=p_2a^2=-\frac{1}{3}, \nonumber \\ &p_3a^2=\frac{2}{3},
\end{eqnarray}
whose average yields the equation of state for dust matter,
$p=0$, as expected.

Hence, the evolution of this universe since the initial,
radiation dominated epoch until the matter dominated one is
similar to that predicted by the open isotropic model, except
for an anisotropy in the pressure distribution, anisotropy that
is negligible at early times and that, for large times, is as
small as the pressures themselves.

However, there is an important difference with respect to the
isotropic case. In the matter dominated era, the energy density
falls as $a^{-3}$, while the pressures decrease as $a^{-2}$. So,
for large times, the magnitude of the pressures would become
larger than the energy density and, consequently, the dominant
energy conditions $\epsilon\geq|p_i|$ would not be fulfilled. It
is possible to prove that, for the energy conditions to be
satisfied at present, the relative energy density should be
larger than or equal to $0.4$, but, even so, these conditions
would be violated sooner or later in the future.

Therefore, in this anisotropic scenario, the dust era should be
followed by an epoch in which the energy density falls, at
least, so slowly as $a^{-2}$, that is, according to the
conservation law
\begin{equation}
\label{quintessencia}
\epsilon a^2=3b^2-1={\rm constant},
\end{equation}
where $b$ is a positive constant introduced for mathematical
convenience (the possibility that $b$ be negative would
correspond to a contracting universe and will not be studied
here).

Substituting this conservation law into Eq. (\ref{E'0}) and
dismissing the terms in $l$ gives $\dot{a}/a=b$, which leads to
the solution
\begin{equation}
\label{quintessencia2}
a=e^{b\eta}=bt.
\end{equation}

The Hubble parameter is now given by $H=b/a$ and, for the
relative energy density, we obtain the constant value
\begin{equation}
\label{omega}
\Omega\equiv\frac{\epsilon}{3H^2}=\frac{3b^2-1}{3b^2}.
\end{equation}

The spatial Einstein equations give
\begin{eqnarray}
&p_1a^2=p_2a^2=-b^2, \nonumber \\ &p_3a^2=1-b^2.
\end{eqnarray}
For the average pressure, we then get $p=-\epsilon/3$, an
equation of state corresponding to a (decaying) positive
cosmological term. In this sense, it is interesting to note
that, for a cosmological term varying with time, the
conservation law $\epsilon a^2={\rm constant}$ has also been
suggested on the basis of quantum cosmology considerations
\cite{Wu}.

It is easy to see that the dominant energy conditions
$\epsilon\geq|p_i|$ are now fulfilled provided that
$b^2\geq1/2$. From (\ref{omega}), one can see that this
corresponds to the condition $\Omega\geq 1/3\approx 0.3$. On the
other hand, it is possible to check that, during the radiation
and matter dominated epochs, the relative energy density
decreases monotonically, which means that the bound obtained
above is a lower limit for $\Omega$ at all times. So, the energy
conditions impose a lower bound on the relative energy density,
maintaining the Universe in a quasi-flat configuration.

The conservation law $\epsilon a^2={\rm constant}$ is also
compatible with Einstein equations in the isotropic case. For
the open FLRW metric, instead of Eq. (\ref{E'0}), we have the
Friedmann equation \cite{Landau}
\begin{equation}
\label{FRW}
\epsilon=\frac{3}{a^4}(\dot{a}^2-a^2).
\end{equation}
Substituting $\epsilon a^2=3(b^2-1)$, it is easy to arrive once
more at the evolution law (\ref{quintessencia2}). In addition,
the spatial Einstein equations give $p=-\epsilon/3$,
corresponding again to a (decaying) positive cosmological term.
In this case, however, although the energy conditions are
satisfied only if $b\geq1$, no positive lower bound is imposed
on the relative energy density, contrary to what happens in the
anisotropic case.

More generally, it is possible to prove that the conservation
law demanding that $\epsilon a^2$ be a constant leads to Eq.
(\ref{quintessencia2}) for all (open, flat and closed) metrics,
in both isotropic and anisotropic cases (actually, in the flat
case the anisotropic metric reduces to the isotropic one).
Remarkably, however, it is only in the open anisotropic case
that the energy conditions impose a quasi-flat configuration.

Let us also note that the anisotropic model presented in this
paper does not exclude the possibility of an inflationary phase
in the cosmic evolution. Indeed, if we add a dominant, positive
cosmological constant to the left hand side of Eq. (\ref{E'0}),
we obtain an exponential evolution law for $a(t)$. Actually, the
introduction of a typical cosmological constant (or,
alternatively, the introduction of an energy density falling
more slowly than $a^{-2}$) in Einstein equations would be needed
if recent claims about the observation of a positive cosmic
acceleration were confirmed \cite{Perlmutter,Riess}. As we have
shown, a cosmological term decaying as $a^{-2}$ leads to the
solution (\ref{quintessencia2}), for which the deceleration
parameter is exactly zero.

\section{The matter content as an imperfect fluid}

As we have seen in section $3$, in the limit of small times the
matter content of space-time (\ref{KO}) consists of rotating
relativistic matter, with energy density and isotropic pressure
given by (\ref{radiation}) and (\ref{pressoes}), respectively.
For large times $l\rightarrow 0$, and from
(\ref{E'1})-(\ref{E'3}) we have
\begin{equation}
\label{l=0}
p_1a^2=p_2a^2=p_3a^2-1.
\end{equation}

In this way, we have an anisotropic distribution of pressures,
with the anisotropy corrections falling as $a^{-2}$. It is this
fact that ultimately leads to the necessity of considering a
coasting evolution (i.e., $a\propto t$) in the last phase of
Universe history, if we want to respect the energy conditions in
a rotating and expanding context.

The appearance of an anisotropic distribution of pressures and,
in particular, of tensions, shows that the material content of
our rotating and expanding universe cannot be a perfect fluid,
as usual in the standard isotropic cosmologies. Therefore, it is
necessary to find a suitable matter source compatible with such
an anisotropy in order to put this non-stationary, rotating
cosmology on an acceptable physical basis.

Without closing the door to other possibilities \cite{CM}, a
natural candidate for the material content is an homogeneous
imperfect fluid with viscosity. Actually, in general situations
this choice is more realistic than a perfect fluid. The use of a
perfect fluid in standard cosmology is possible due to the
isotropy of the fluid motion, which avoids the appearance of
friction. But in anisotropic contexts like the one considered
here the role of viscosity cannot, in general, be neglected.

In order to simplify our analysis, let us consider the fluid
motion in a locally comoving Lorentz frame, that is, a freely
falling frame whose origin is at rest with respect to a given
point of the fluid at a given time $t$ (owing to the presence of
friction, we cannot introduce a globally comoving frame). The
energy-momentum tensor of the fluid is given by
\cite{Ibidem,Misner}
\begin{equation}
\label{T}
T^{\mu\nu}=-pg^{\mu\nu}+(p+\epsilon)U^{\mu}U^{\nu}+\Delta
T^{\mu\nu},
\end{equation}
where $p$ is the average pressure and, in our frame, $\Delta
T^{00}=0$ and
\begin{eqnarray}
\label{Deltas}
&\Delta T^{i0}=-\chi_0(\partial^i T + T\dot{U}^i),\nonumber\\
&\Delta T^{ij}=-\eta_0(\partial^j U^i + \partial^i U^j
-\frac{2}{3} \partial_k U^k \delta^{ij}).
\end{eqnarray}
Here, $U^{\mu}$ is the fluid velocity, $T$ its temperature, the
overdot means derivation with respect to $t$ and
$\partial^i=\delta^{ij}\partial_j$. $\chi_0$ and $\eta_0$ are
the fluid heat conductivity and shear viscosity, respectively.

In our model with cylindrical symmetry, the second of equations
(\ref{Deltas}) leads to
\begin{eqnarray}
\label{Deltas2}
&\Delta T^{11}=\Delta T^{22}=-\frac{2}{3}\eta_0h, \nonumber \\
&\Delta T^{33}=\frac{4}{3}\eta_0h,
\end{eqnarray}
where we have defined $h\equiv\partial^1 U^1 - \partial^3 U^3 =
\partial^2 U^2 - \partial^3 U^3$. So, we have
\begin{equation}
\label{Deltas3}
\Delta T^{11}=\Delta T^{22}=\Delta T^{33}-2\eta_0h.
\end{equation}

Comparing this equation with (\ref{l=0}), we see that viscosity
gives the expected contribution to the pressures provided that
\begin{equation}
\label{h}
2\eta_0ha^2=1.
\end{equation}

Since the anisotropic metric (\ref{aniso}) is diagonal, the
energy-momentum tensor will also be diagonal, that is,
$T^{i0}=\Delta T^{i0}=0$. This can be achieved if we consider a
null heat conductivity, or an approximately null fluid
temperature, or yet if the condition $\partial^i T=-T\dot{U}^i$
follows. In this last case, since for the coasting solution
considered in the last section we have $\dot{U}^i=0$, we get
$\partial_i T=0$, which means isotropy and homogeneity of fluid
temperature.

On the other hand, we have $\Delta T^{ij}=0$ for $i\neq j$,
which leads to $\partial^j U^i=-\partial^i U^j$ for $i\neq j$.
This condition can be satisfied, in particular, by the Hubble
type law $U_i=H_ix_i$, $i=1,2,3$. In this case, equation
(\ref{h}) shows that $H_3<H_1=H_2$, and it is important to
verify how much this inequality can affect the observed Hubble
law.

The relation between the Hubble parameter $H$ and the parameters
$H_i$ can be established with the help of the equation
\cite{Misner} $\dot{V}/V=\partial_i U^i$, relating the temporal
variation of a volume $V$ of the fluid and the divergence of its
velocity field. Considering $V$ as the volume of the observed
universe, proportional to $a^3$, this leads to $\partial_i
U^i=3\dot{a}/a=3H$. So, from $U_i=H_ix_i$, we obtain
$H=(H_1+H_2+H_3)/3$, that is, the Hubble parameter is equal to
the average of the paramaters $H_i$.

During the coasting phase of Universe evolution, the Hubble
parameter is given by $H=b/a$ and, therefore, its relative
variation with direction is given by $h/H=(2\eta_0 ba)^{-1}$. If
we consider a constant viscosity $\eta_0$ (or at least a
viscosity varying very slowly with $a$), the relative variation
of $H$ falls with $a$ and could be unobservable for large times,
even for a small value of $\eta_0$.

Therefore, the above analysis shows that the solutions found for
the spacetime metrics (\ref{KO}) and (\ref{aniso}) may
correspond to a physical matter content formed by radiation,
viscous matter and a cosmological term. The appearance of
anisotropic pressures, far from discard the model and render it
unphysical, can be related to the presence of friction owing to
the anisotropic motion of matter.

\section{A singularity-free alternative scenario}

In a recent paper \cite{PRD}, we have investigated an
alternative, singularity free, scenario for Universe's
evolution, in which the present expanding universe is originated
from a primordial G\"{o}del universe \cite{Godel}, by a phase
transition during which the negative cosmological term
characteristic of the G\"{o}del phase crosses a positive maximum and
rolls down to zero. This scenario could also explain the origin
of galaxies rotation, but it was not clear how the global
angular momentum of the Universe could be transferred to the
galaxies \cite{Mathews}. This difficult is intimately connected
to the discontinuous transition considered in that paper, where
the G\"{o}del metric (\ref{KO}) (with $l=\sqrt{2}$) is directly
matched with the expanding (but non-rotating) anisotropic metric
(\ref{aniso}).

The analysis we have made in the present paper can help us
solving these difficulties. Initially we have a G\"{o}del universe,
which corresponds to $l=\sqrt{2}$ and $\epsilon a^2=1$. After
the phase transition, we have a rotating and expanding universe,
in which $l$ (and then the non-diagonal term in metric
(\ref{KO})) falls down, leading to the diagonal metric
(\ref{aniso}). During the phase transition, half of the initial
energy is used to compensate the negative cosmological constant
present in the G\"{o}del model, given by $\lambda a^2=-1/2$. After
the phase transition, we then have $\epsilon a^2=1/2$, relation
which allows us to match the original G\"{o}del universe with our
last solution (\ref{quintessencia})-(\ref{quintessencia2}) for
$b^2=1/2$, that is, for $\Omega\approx 0.3$. In this way, the
scale factor $a$ changes continuously in the whole evolution,
and the dominant energy conditions are satisfied. Moreover, the
decaying, positive cosmological term characteristic of the
expanding phase can be shown to arise naturally from the scalar
field transition described in \cite{PRD}, in which a
self-interaction potential, initially at a negative minimum
(corresponding to the negative cosmological constant present in
the G\"{o}del solution), crosses a positive maximum and rolls down
to zero. Now, we can use the mechanism proposed by Li to
transfer angular momentum from Universe to galaxies in a
rotating and expanding context \cite{Li}.

In the above match, the value of the radius of the primordial
universe is not fixed by the present values of the energy
density and Hubble parameter, contrary to what occurs in the
match considered in Ref. \cite{PRD} (there, the G\"{o}del phase is
matched with the dust solution given by (\ref{intermediate}) and
(\ref{intermediate2})). In the G\"{o}del phase we have the angular
velocity $\omega_G=\sqrt{2}/2a_G$, where $a_G$ is the radius of
the primordial G\"{o}del's universe. In the expanding phase, the
angular velocity of matter at the present time is given by
$\omega a^2=\omega_G a_G^2$. Substituting this last equation
into the former, we obtain $a_G=\sqrt{2}\omega a^2$, which
determines $a_G$ for a given value of the present angular
velocity of matter.

As already discussed in \cite{PRD}, in this context there is no
dense phase, which constitutes the major drawback of this
alternative scenario, if we have into consideration phenomena
like nucleosynthesis or the cosmic background radiation.
Although such phenomena could be related to the very process of
phase transition \cite{PRD}, this possibility needs to be
further investigated.

\section{Concluding remarks}

In this paper we have tried to show that the inclusion of
rotation into the standard model of the Universe can enrich it
in several aspects. On one hand, rotation does not contradict
the current observations of isotropy nor gives rise to parallax
effects \cite{KO,Obukhov}; the anisotropic metric (\ref{aniso}),
that we have just shown to originate from the rotating metric
(\ref{KO}) by conservation of angular momentum, reduces to the
open metric of Friedman in the limit of nearby distances; in the
limit of small times the distribution of pressures is isotropic,
and the smallness of the rotation parameter $l$ guarantees that
physical processes taking place at early times are not affected
by rotation, as well as the absence of closed time-like curves.

On the other hand, the global rotation can be used to explain
the origin of galaxies rotation and the observed relation
between their angular momenta and masses \cite{Li}; in the
anisotropic context described by metrics (\ref{KO}) and
(\ref{aniso}), the energy conditions lead naturally to a last
epoch dominated by a positive cosmological term decaying as
$a^{-2}$, a decaying law also expected on the basis of quantum
cosmology reasonings \cite{Wu}; finally, in such a context, the
energy conditions impose also a constant lower bound on the
relative energy density, close to the present observed value,
providing in this way a possible explanation for the observed
quasi-flatness. By the way, let us note that these two last
results are originated from the anisotropy of metric
(\ref{aniso}), no matter its relation with the rotating metric
(\ref{KO}).

As recently pointed out \cite{JJ}, the coasting evolution law
$a=bt$, characteristic of the last phase of the present model,
can solve other cosmological problems as well. For example, it
leads to $t=1/H$, an age for the Universe compatible with the
observational bounds. In addition, the conservation law
$\epsilon a^2=$ constant is precisely what we need to solve the
cosmological constant problem, obtaining a cosmological term in
agreement with observation \cite{Wu}. It has also been shown
\cite{JJ} that the decaying cosmological term proposed by Chen
and Wu is not the only feasible possibility, the equation of
state $p=-\epsilon/3$ being also compatible with a bicomponent
content, formed by ordinary matter and a cosmological constant.
Actually, this equation of state can correspond as well to
textures or strings. However, as commented in Ref. \cite{JJ}, it
would be unrealistic to consider that the present universe is
dominated by such topological defects.

Finally, a curious remark is in order. With the superior limit
for the matter angular velocity $\omega\sim10^{-21}s^{-1}$,
derived in section 2, the radius of the Universe
$a\sim10^{26}$m, and the matter density
$\rho\sim10^{-27}$Kg/m$^3$, we obtain an angular momentum of
order $L\sim10^{82}$J.s. With this value, it follows that
\begin{equation}
\frac{L}{\hbar}\sim10^{116}\sim(10^{39})^3,
\end{equation}
where $\hbar$ is Planck's constant. This relation between the
angular momentum of the Universe and typical angular momenta of
particles is also expected on the basis of the large number
coincidences \cite{FPL,Dirac}. Its physical meaning and
cosmological implications remain an exciting and challenging
open problem.

\section*{Acknowledgments}

I am greatly thankful to Guillermo A. Mena Marug\'{a}n, without
whose criticisms, suggestions and encouragement this work would
have not been possible. I am also grateful to Luis J. Garay for
his careful reading of a first version of the manuscript, to
Mariano Moles for a useful discussion on galaxies formation, to
Hugo Reis and Henrique Fleming for discussions on the Hubble law
in anisotropic models, to Pedro F. Gonz\'{a}lez-D\'{\i}az for interesting
discussions and warm hospitality in CSIC, and to Enric Verdaguer
and Alberto Saa for their hospitality in University of Barcelona
during the Summer of 2000. My thanks also to Mariam Bouhmadi for
her kind help with GRTensor and Mathematica. This work has been
partially supported by CNPq.

\end{document}